\documentclass[10pt,final,journal,twocolumn]{IEEEtran}
\ifCLASSINFOpdf
\else
\fi
\usepackage{cite}
\usepackage[cmex10]{amsmath}
\usepackage{amsthm}
\usepackage{algorithmic}
\usepackage{stfloats}
\usepackage{multicol,multienum}
\usepackage[dvips]{graphicx}
\begin{document}

\title{ Fast Load-aware Spectrum Allocation for Cognitive Small Cells: A Cloud-based Game-theoretic Solution}

\title{Database-assisted Spectrum Access in Dynamic Networks: A  Distributed Learning Solution }


\author{Yuhua~Xu,~\IEEEmembership{Member,~IEEE,}
Yitao Xu
and Alagan~Anpalagan,~\IEEEmembership{Senior Member,~IEEE}
\thanks{This work was supported by the National Science Foundation of China
under Grant No. 61401508 and No. 61172062. }

%
\thanks{ Yuhua Xu and Yitao Xu  are with the College of Communications Engineering, PLA University of Science and Technology, Nanjing, 21007, China. (e-mail: yuhuaenator@gmail.com; yitaoxu@126.com) }
%
\thanks{ A.~Anpalagan is with the Department of Electrical and Computer Engineering, Ryerson University, Toronto,
 Canada (e-mail: alagan@ee.ryerson.ca).}
}

%
%

\IEEEpeerreviewmaketitle
\maketitle

\begin{abstract}
This paper investigates the problem of database-assisted spectrum access in dynamic TV white spectrum networks, in which the active user set is varying. Since there is no central controller and information exchange, it encounters \emph{dynamic} and \emph{incomplete} information constraints.
To solve this challenge, we formulate a state-based spectrum access game and a robust spectrum access game.
It is proved that the two games are ordinal potential games with the (expected)  aggregate weighted interference serving as the potential functions. A distributed learning algorithm  is proposed to achieve the pure strategy Nash equilibrium (NE) of the games. It is shown that the best NE is almost the same with the optimal solution and the achievable throughput of the proposed learning algorithm is very close to the optimal one, which  validates the effectiveness of the proposed game-theoretic solution.

\end{abstract}

\begin{IEEEkeywords}
  TV white spectrum, geo-location database, ordinal potential game,   learning automata.
\end{IEEEkeywords}

%
\IEEEpeerreviewmaketitle

\section{Introduction}
\IEEEPARstart {E}{mploying} TV White Spectrum (TVWS) \cite{TVWS1,TVWS2,TVWS3} has been regarded as a promising approach to solve the spectrum shortage problem in future wireless networks, as it can effectively improve the spectrum efficiency by allowing the unlicensed users dynamically to access the idle TV channels. For the TVWS, it has been shown that obtaining the spectrum availability information by inquiring a geo-location database is more efficient than performing spectrum sensing alone \cite{database1,database2}.
Also, geo-location  database  has been widely supported by the standards bodies and industrial organizations \cite{database_DySPAN,database_google,database_alliance,database_microsoft}.

\begin{figure*}[!tb]
\centering
\includegraphics[width=4.8in]{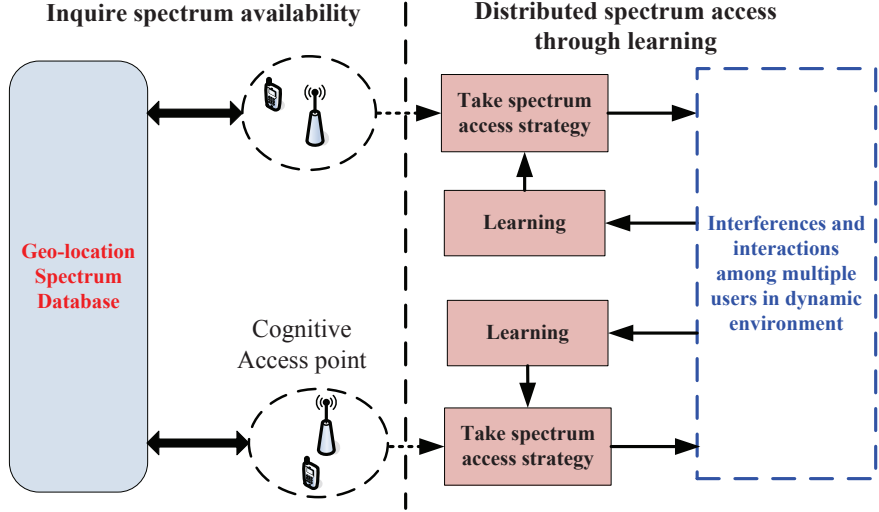}
\caption{The illustrative diagram of database-assisted spectrum access.  }
\label{fig:system_model}
\end{figure*}

Currently, researchers in this field mainly focused on: i) constructing and maintaining the geo-location database, e.g., \cite{database_maintain,database_radiomap,database_indoor}, and ii) developing business models to analyze the revenues of the TV spectrum holders and the unlicensed users, e.g., \cite{database_business1,database_business2,database_business3}.
Since there is no centralized controller available, how to choose a channel for transmission in a distributed manner, aiming to avoid mutual interference among the users, remains a key challenge. However, only a few preliminary results  were reported recently \cite{database_access1,database_access2}, and hence it is urgent and important to study efficient database-assisted distributed spectrum access strategies.

In this paper, we consider a dynamic and distributed TVWS network. Specifically,  considering practical applications of the users, they do not access the channels when there is no data to transmit. To capture this dynamics, it is assumed that each user becomes active/inactive according to an active probability in each decision period. As a result, the active user set is varying. Furthermore, since there is no information exchange, a user does not know the chosen channels, the current state (active or inactive) and the active probabilities of other users. The \emph{dynamic} and \emph{incomplete} information constraints make the task of developing efficient distributed spectrum access strategies challenging.

To solve this problem, we resort to game models \cite{game_book} and  distributed learning technology. Specifically, we  formulate a state-based spectrum access game and a robust spectrum access game, and propose a distributed learning to achieve desirable solutions. The main contributions can be summarized as follows:
\begin{enumerate}
  \item For an arbitrary active user set, we formulate the problem of distributed spectrum access as a state-based non-cooperative game. It is proved that state-based game is an ordinal potential game with the aggregate weighted interference serving as the potential function. To address the challenges caused by the varying active user set, we formulate a robust spectrum access game, which is also proved to be a ordinal potential game. Finally, we propose a distributed learning algorithm to achieve the pure strategy Nash equilibria of the formulated games.
  \item It is shown that the best Nash equilibrium is almost the same with optimal solution, which validates the effectiveness of the formulated game models. In addition, the achievable throughput  of the proposed learning algorithm is very close to the optimal one, which also validates the distributed learning algorithm in dynamic networks.
\end{enumerate}

The most related work is \cite{database_access1}, in which  game-theoretic database-assisted spectrum sharing strategies were investigated. This work is differentiated in: i)
all users are assumed to be always active \cite{database_access1}, while we consider a network with varying number of active users, and iii) the spectrum access algorithms in \cite{database_access1} need information exchange, while the proposed learning-based spectrum access algorithm is fully distributed. Also, the problem of distributed spectrum access for minimizing the aggregate weighted interference was studied in \cite{weighted_interference1} and in our previous work \cite{weighted_interference2,weighted_interference3}. The differences in the current work are that we optimize the throughput directly and the active user set in each decision period is randomly changing. Also, some  simulation results on the problem of distributed spectrum access with changing active user set were preliminarily reported in our recent work \cite{Xu_magazine}. In this work, rigorous analysis and proofs are given and more simulation results are presented to validate the proposed game-theoretic solution.

The rest of this paper is organized as follows. In Section II, the system model and problem formulation are presented. In Section III, we formulate the state-based spectrum access game and the robust spectrum game respectively, analyze their properties, and propose a distributed learning-based spectrum access algorithm. Finally, simulation results and discussion are presented in Section IV and conclusion is drawn in Section V.

\section{System Model and Problem Formulation}
\subsection{System model}

We consider a distributed network with $N$ cognitive users and $M$ channels. Note that here each cognitive user
corresponds to a communication link consisting of a transmitter and a receiver, e.g., the cognitive access point (AP) and its serving clients. Each cognitive user inquires the spectrum availability from the geo-location spectrum database, which specifies the available channel set $\mathcal{A}_n$ and the maximum allowable transmission power $P_n$ for each user $n$. With the inquired information on the available channels  and maximum transmission power, the users take suitable spectrum access strategies through learning. An illustration of the database-assisted spectrum access is shown in Fig. \ref{fig:system_model}.

To address the user traffic in practical applications, we consider a  network with a varying number of active users.
 Specifically, it is assumed that each user performs channel access in each slot with probability $\lambda_n$, $0 < \lambda_n \le 1$. Such a model captures general kinds of dynamics in wireless networks \cite{Xu_TVT2015}, e.g., a user becomes active only when it has data to transmit and inactive otherwise; also, it can be regarded as a high-level abstraction of the user traffic, i.e., the active probability corresponds to the probability of non-empty buffer.

 \subsection{Problem formulation}
  To capture the changing number of active users, we define the system state as $\mathbf{S}=\{s_1, \ldots, s_N\}$, where  $s_n=1$ indicates that user $n$  is active while $s_n=0$ means  it is inactive. The  system state probability is then given by $\mu(s_1, \ldots, s_N)=\prod\nolimits_{n=1}^{N} p_n$, where $p_n$ is determined as follows:
  \begin{equation}
    p_n=\left\{ \begin{array}{l}
 \lambda_n,\;\;\;\;\;\;\;\;\;\;\;s_n=1 \\
 1-\lambda_n,\;\;\;\;\; s_n=0\\
 \end{array} \right.
 \end{equation}

 Denote an arbitrary and non-empty active user set as $\mathcal{B}$, i.e., $\mathcal{B}=\{n\in\mathcal{N}:s_n=1\}$. For presentation, denote the set of all the active user sets as $\Gamma$. Then, the probability of an active user set can denoted by $\mu(\mathcal{B})$. We have $\sum\nolimits_{\mathcal{B}\in\Gamma}\mu(\mathcal{B})=1-\mu(\mathcal{B}_0)$, where $\mu(\mathcal{B}_0)= \prod\nolimits_{n=1}^{N} (1-\lambda_n)$ is the probability that all the users are inactive.

 With the spectrum availability information obtained from the  database,  user $n$ chooses a channel $a_n \in \mathcal{A}_n$ for data transmission. For any active user set $\mathcal{B}$ and  channel selection profile $(a_n, a_{-n})$, the received Signal-to-Interference-plus-Noise Ratio (SINR) of an active user $n$ is determined by:
  \begin{equation}
    \eta_n(\mathcal{B},a_n,a_{-n})=
 \frac{{P_{n}}d_{n}^{-\alpha }}{{\sum\nolimits_{i\in \mathcal{B}\backslash \{n\}:{{a}_{i}}={{a}_{n}}}}{{P}_{i}}d_{in}^{-\alpha }+\sigma},
 \end{equation}
where $\alpha$  is the path loss factor, $d_n$ is the distance between user $n$ and its dedicated receiver, $d_{in}$ is the distance between user $i$ and $n$, ${\sum\nolimits_{i\in \mathcal{B}\backslash \{n\}:{{a}_{i}}={{a}_{n}}}}{{P}_{i}}d_{in}^{-\alpha }+\sigma$ is the aggregate interference from all other active users also choosing channel $a_n$, and $\sigma$ is the background noise. Then, the achievable throughput of user  $n$ is given by:
  \begin{equation}
    R_n(\mathcal{B},a_n,a_{-n})=B \log \big(1+\eta_n(\mathcal{B},a_n,a_{-n})\big),
\label{eq:instantaneous_payoff}
 \end{equation}
where $B$ is the channel bandwidth. Therefore, there are two possible optimization goals for each user, i.e.,
\begin{equation}
\label{eq:Goal1}
\textbf{P1}: \;\;\;\;\mathop {\max }\limits_{a_n} \; R_n(\mathcal{B},a_n,a_{-n}), \forall \mathcal{B}\in \Gamma,
\end{equation}
or
\begin{equation}
\label{eq:expected_network_throughput}
\textbf{P2}: \;\;\;\;\mathop {\max }\limits_{a_n} \; \text{E}_{\mathcal{B}} [R_n(\mathcal{B},a_n,a_{-n})]=\sum\limits_{\mathcal{B} \in \Gamma } \mu(\mathcal{B})R_n(\mathcal{B},a_n,a_{-n})
\end{equation}

Since the network is always changing, it is not feasible to optimize the achievable throughput for each active user set. Thus, we pay attention  to solving   $\textbf{P2}$. However, the task of solving $\textbf{P2}$ is challenging due to the following imperfect information constraints:
\begin{itemize}
  \item \textbf{Dynamic:} the active user set in the system is always changing; in particular, it may change randomly in each decision period.
  \item \textbf{Incomplete:} there is no information exchange among the users, which leads to: i) a user does not know the active probabilities and chosen channels  of other users, and ii) the  state distribution $\mu(\mathcal{B})$ is unknown to the all users.
\end{itemize}

Based on the above analysis, it is seen that centralized approaches are not available and we need to  develop a distributed and learning-based approach for solving problem $\textbf{P2}$.





\section{ Spectrum Access Games and Distributed Learning Algorithm }
Since no centralized controller is  available in the considered distributed network and all the users take their spectrum access strategies distributively and autonomously, we formulate this problem as a non-cooperative game. In the following, we present the formulated game models, analyze its properties, and propose a distributed learning algorithm to converge to stable solutions in dynamic environment.

\subsection{Game formulation and analysis}
In this subsection, we first present a state-based spectrum access game, in which an inherent system state specifies the active user set.  Based on the state-based game, we then present a robust game, in which the expectations over all possible system states are considered. Note that the state-based game corresponds to problem $\textbf{P1}$ while the robust game corresponds to problem $\textbf{P2}$.

\subsubsection{State-based spectrum access game}
Formally, the state-based spectrum access game model is denoted as $\mathcal{F}_1=[\mathcal{N}, \mathcal{B},{\{{\mathcal{A}_n}\}_{n\in\mathcal{N}}},{\{{u_n(\mathcal{B})}\}_{n\in\mathcal{N}}}]$, where  $\mathcal{N} = \{1,\ldots,N\}$ is a set of players (users), $\mathcal{B}$ is active user set, $\mathcal{A}_n$ is a set of  available actions (channels) for user $n$,  and $u_n(\mathcal{B})$ is the utility function of player $n$.  The utility function is defined as the available transmission, i.e.,
\begin{equation}
\label{eq:utility_function}
u_n(\mathcal{B},a_n,a_{-n})=R_n(\mathcal{B},a_n,a_{-n}), \forall n \in \mathcal{N}, \forall \mathcal{B} \in \Gamma
\end{equation}
Each user is the game intends to maximize its individual utility, which means that the state-based spectrum access game can be expressed as:
\begin{equation}
\label{eq:game_model1}
(\mathcal{F}_1): \;\;\;\;\;\;\;\; \mathop {\max }\limits_{a_n \in A_n} u_n(\mathcal{B},a_n,a_{-n}),\forall n \in \mathcal{B}.
\end{equation}


In order to investigate the properties of $\mathcal{F}_1$, we first present the following  definitions, which is directly drawn from \cite{Monderer96}.
\\
\textbf{Definition 1 (Nash equilibrium )}. For an arbitrary active user set $\mathcal{B}$, a spectrum access  profile $a^*=(a^*_1,\ldots,a^*_{|\mathcal{B}|})$ is a pure strategy NE  if and only if no player can improve its utility  by deviating unilaterally, i.e.,
\begin{equation}
\label{eq:NE_definition}
 {u_n}(\mathcal{B},{a^*_n},a^*_{-n}) \ge  {u_n}(\mathcal{B},a_n,{a^*_{-n}}), \forall n \in \mathcal{B}, \forall a_n \in \mathcal{A}_n, a_n\ne a^*_n
\end{equation}
\textbf{Definition 2 (Ordinal potential game  )}. A game is an ordinal potential game (OPG) if there exists an ordinal potential function $\phi: {{\mathcal{A}}_1} \times  \cdots  \times {{\mathcal{A}}_N} \to R$ such that for all $n \in \mathcal{N}$, all $a_n \in \mathcal{A}_n$, and $a'_n \in \mathcal{A}_n$, the following holds:
 \begin{equation}
 \label{eq:OPG_definition}
  \begin{array}{l}
    u_n(\mathcal{B},a_n,a_{-n})-u_n(\mathcal{B},a'_n,a_{-n}) >0 \\
    \;\;\;\;\;\;\;\;\;\;\;\;\;\;\;\;\Leftrightarrow \phi(\mathcal{B},a_n,a_{-n})-\phi(\mathcal{B},a'_n,a_{-n})>0
  \end{array}
 \end{equation}
That is, the change in the utility function caused by  the unilateral action change of an arbitrary each user has the same trend with that in the ordinal potential function. Following the similar methodology presented in \cite{database_access1}, we have the following theorem.

\newtheorem{theorem}{Theorem}
\begin{theorem}
\label{tm:state_potential_game}
For any active user set $\mathcal{B}$, the state-based spectrum access game $\mathcal{F}_1$ is an ordinal potential game.
\end{theorem}

\begin{IEEEproof}
For presentation, for any active user set $\mathcal{B}$ and an arbitrary user $ \forall n \in \mathcal{B}$, denote
 \begin{equation}
    v_n(\mathcal{B},a_n,a_{-n})=-{\sum\limits_{i\in \mathcal{B}\backslash \{n\}:{{a}_{i}}={{a}_{n}}}} {{P}_{i} P_n}d_{in}^{-\alpha },
 \end{equation}
which can be regarded as the weighted interference \cite{weighted_interference1} experienced by user $n$. Define  $\phi: {{\mathcal{A}}_1} \times  \cdots  \times {{\mathcal{A}}_{|\mathcal{B}|}} \to R$ as
 \begin{equation}
\label{eq:state_potential}
\begin{array}{l}
    \phi(\mathcal{B},a_n,a_{-n})= \sum\limits_{n \in \mathcal{B}} v_n(\mathcal{B},a_n,a_{-n})\\
\;\;\;\;\;\;\;\;\;\;\;\;\;\;\;\;\;\;\;\;\;\;\;=-\sum\limits_{n \in \mathcal{B}} {\sum\limits_{i\in \mathcal{B}\backslash \{n\}:{{a}_{i}}={{a}_{n}}}}{{P}_{i} P_n}d_{in}^{-\alpha },
\end{array}
 \end{equation}
which is the aggregate weighted interference experienced by all the active users.

If an arbitrary player $n$ unilaterally changes its channel selection from $a_n$ to $a^*_n$, then the change in
 $v_n$ caused by  this unilateral change is as follows:
 \begin{equation}
\begin{array}{l}
v_n(\mathcal{B},a^*_n,a_{-n})-v_n(\mathcal{B},a_n,a_{-n})\\
={\sum\limits_{i\in \mathcal{B}\backslash \{n\}:{{a}_{i}}={{a}_{n}}}} {{P}_{i} P_n}d_{in}^{-\alpha }-{\sum\limits_{i\in \mathcal{B}\backslash \{n\}:{{a}_{i}}={{a}^*_{n}}}} {{P}_{i} P_n}d_{in}^{-\alpha }
\end{array}
 \end{equation}

For analysis convenience, denote the users choosing the same channel with user $n$ as $\mathcal{I}_n(a_n, \mathcal{B})=\{i \in \mathcal{B} \backslash n: a_i=a_n\}$. Then, the change in potential function $\phi$ caused by the unilateral change of user $n$ can be expressed as follows:
 \begin{equation}
\label{eq:potential_challenge1}
\begin{array}{l}
\phi(\mathcal{B},a^*_n,a_{-n})-\phi(\mathcal{B},a_n,a_{-n})\\
=v_n(\mathcal{B},a^*_n,a_{-n})-v_n(\mathcal{B},a_n,a_{-n})\\
\;\;+{\sum\limits_{i \in \mathcal{I}_n(a^*_n, \mathcal{B})} {{P}_{n} P_i}d_{ni}^{-\alpha }
-{\sum\limits_{i \in \mathcal{I}_n(a_n, \mathcal{B})}}} {{P}_{n} P_i}d_{ni}^{-\alpha }\\
=2 \Big(v_n(\mathcal{B},a^*_n,a_{-n})-v_n(\mathcal{B},a_n,a_{-n})\Big)
\end{array}
 \end{equation}

Since $u_n$ and $v_n$ is related by:
  \begin{equation}
    u_n(\mathcal{B},a_n,a_{-n})=
B \log \Big(1+ \frac{{P_{n}}d_{n}^{-\alpha }}{-v_n(\mathcal{B},a_n,a_{-n})/P_n+\sigma}\Big),
 \end{equation}
and   $\log \Big(1+\frac{{P_{n}}d_{n}^{-\alpha }}{-x/P_n+\sigma} \Big)$ is increasing with respect to $x$, it follows that:
  \begin{equation}
\label{eq:potential_challenge2}
\begin{array}{l}
    \Big(u_n(\mathcal{B},a^*_n,a_{-n})-u_n(\mathcal{B},a_n,a_{-n})\Big)\\
\times \Big(v_n(\mathcal{B},a^*_n,a_{-n})-v_n(\mathcal{B},a_n,a_{-n})\Big) \ge 0, \forall a_n,a^*_n \in \mathcal{A}_n
\end{array}
 \end{equation}
For any active user set $\mathcal{B}$, combining (\ref{eq:potential_challenge1}) and (\ref{eq:potential_challenge2}) yields the following:
  \begin{equation}
\label{eq:potential_challenge3}
\begin{array}{l}
    \Big(u_n(\mathcal{B},a^*_n,a_{-n})-u_n(\mathcal{B},a_n,a_{-n})\Big)\\
\times \Big(\phi(\mathcal{B},a^*_n,a_{-n})-\phi(\mathcal{B},a_n,a_{-n})\Big) \ge 0, \forall a_n,a^*_n \in \mathcal{A}_n
\end{array}
 \end{equation}
which  satisfies the definition of OPG, as characterized by  (\ref{eq:OPG_definition}). Thus, the state-based spectrum access game $\mathcal{F}_1$ is an ordinal potential game with $\phi$ serving as the potential function, which proves Theorem \ref{tm:state_potential_game}.
\end{IEEEproof}

\subsubsection{Robust spectrum access game} As discussed above, it is not feasible to perform optimization for each active user set since the network is always changing. Thus, based on the state-based spectrum access game $\mathcal{F}_1$, we formulate a robust spectrum game below. Specifically, the robust spectrum access game is denoted as $\mathcal{F}_2=[\mathcal{N}, {\{{\mathcal{A}_n}\}_{n\in\mathcal{N}}},{\{\omega_n\}_{n\in\mathcal{N}}}]$, where  $\mathcal{N} = \{1,\ldots,N\}$ is a set of players (users), $\mathcal{A}_n$ is a set of  available actions (channels) for user $n$,  and $\omega_n$ is the utility function of player $n$.  The utility function in robust spectrum access game is defined as the expected available transmission rate over all possible active user sets \cite{Xu_magazine,Xu_TVT2015,Learning_book}, i.e.,
\begin{equation}
\label{eq:Goal1}
\omega_n(a_n,a_{-n})=\text{E}_{\mathcal{B}} [u_n(\mathcal{B},a_n,a_{-n})]=\sum\limits_{\mathcal{B} \in \Gamma } \mu(\mathcal{B})u_n(\mathcal{B},a_n,a_{-n})
\end{equation}
Similarly, the robust spectrum access game can be expressed as:
\begin{equation}
\label{eq:game_model2}
(\mathcal{F}_2): \;\;\;\;\;\;\;\; \mathop {\max }\limits_{a_n \in A_n} \omega_n(a_n,a_{-n}),\forall n \in \mathcal{N}.
\end{equation}


\begin{theorem}
\label{tm:robust_potential_game}
The robust spectrum access game $\mathcal{F}_2$ is also an ordinal potential game.
\end{theorem}
\begin{IEEEproof}
We define the potential function as:
\begin{equation}
\label{eq:robust_potential}
\Phi(a_n, a_{-n})= \text{E}_{\mathcal{B}} \big[\phi(\mathcal{B},a_n, a_{-n})\big],
\end{equation}
where $\phi$ is specified by (\ref{eq:state_potential}).

If an arbitrary player $n$ unilaterally changes its channel selection from $a_n$ to $a^*_n$, then the change in
 $w_n$ caused by  this unilateral change is as follows:
 \begin{equation}
\begin{array}{l}
\omega_n(a^*_n,a_{-n})-\omega_n(a_n,a_{-n})\\
=\text{E}_{\mathcal{B}} [u_n(\mathcal{B},a^*_n,a_{-n})-u_n(\mathcal{B},a_n,a_{-n})]
\end{array}
 \end{equation}

Similarly, the challenge in the $\Phi$ is determined by:
 \begin{equation}
\begin{array}{l}
\Phi(a^*_n,a_{-n})-\Phi(a_n,a_{-n})\\
=\text{E}_{\mathcal{B}} [\phi(\mathcal{B},a^*_n,a_{-n})-\phi(\mathcal{B},a_n,a_{-n})]
\end{array}
 \end{equation}
Using the result obtained in (\ref{eq:potential_challenge3}), the following always holds:
  \begin{equation}
\begin{array}{l}
    \Big(\omega_n(a^*_n,a_{-n})-\omega_n(a_n,a_{-n})\Big)\\
\times \Big(\Phi(a^*_n,a_{-n})-\Phi(a_n,a_{-n})\Big) \ge 0, \forall a_n,a^*_n \in \mathcal{A}_n
\end{array}
 \end{equation}
Thus, it is proved that the robust spectrum access game $\mathcal{F}_2$ is also an ordinal potential game with $\Phi(a_n, a_{-n})$ serving as the potential function.
\end{IEEEproof}

\subsubsection{Discussion on the game models}
Ordinal potential game (OPG) admits the following two promising features \cite{Monderer96}: (i) every OPG has at least one pure strategy Nash equilibrium, and (ii) an action profile that maximizes the ordinal potential function is also a pure strategy Nash equilibrium. Some further discussions on the two spectrum game models are listed below:

\begin{itemize}
  \item Both the state-based and robust spectrum access games have at least one pure strategy NE.
  \item For the state-based spectrum access game, the aggregate weighted interference serves as the potential function, as specified by (\ref{eq:state_potential}). For the robust spectrum access game, the expected aggregate weighted interference serve as the potential game, as specified by (\ref{eq:robust_potential}). Thus, it is known all the NEs of the games minimize the (expected) aggregate weighted interference respectively. Furthermore, it has been shown that minimizing lower weighted interference leads to higher throughput \cite{weighted_interference1,weighted_interference2,weighted_interference3}. Thus, it can be expected that the two games would also achieve high throughput.
\end{itemize}

\subsection{Distributed learning for achieving  Nash equilibria}
As the expected aggregate interference serves as the potential function for the robust spectrum access game, it is desirable to develop distributed algorithms to achieve the Nash equilibria. Conventional algorithms in the game community, e.g., best response dynamic \cite{Monderer96}, fictitious play \cite{fictitious_play} and spatial adaptive play \cite{spatial_adaptive_play}, can not be applied since they need information exchange among the players. To eliminate the requirement of information exchange, some distributed algorithms have been applied in wireless applications, e.g.,  MAX-logit \cite{MAX_logit} and Q-learning \cite{Q_learning}. However, B-logit, and MAX-logit  are only suitable for static environment;  although Q-learning can be applied in dynamic networks, its convergence in multiuser environment can not be guaranteed.

In this paper, we propose a learning-based distributed spectrum access algorithm, which is mainly based the stochastic learning automata \cite{Verbeeck02,Sastry94}. To begin with, denote  $\mathbf{q}_n(k)=\{q_{n1}(k), \ldots, q_{n|\mathcal{A}_n|}(k)\}$ as the mixed strategy of player $n$ in the $k$th slot, in which $q_{nm}$ is the probability of choosing channel $m$. The key ideas of the proposed distributed learning algorithm are: i) the active users choose the channels according to their mixed strategies,  and then update their mixed strategies based on the received payoffs, and iii) an inactive user does nothing. Specifically, the learning procedure is as follows:
\\
 \rule{\linewidth}{1pt}
\\
{\textbf{Initialization}:} set $k=1$ and set the initial mixed strategy of each user as $q_{nm}(k) = \frac{1}{|\mathcal{A}_n|},\forall n \in \mathcal{N},\forall m\in \mathcal{A}_n$.
\\
\textbf{Loop for $k=1,2,\ldots$}

Denote $\mathcal{B}(k)$ as the active user set in the current slot.

 \textbf{1). Channel selection:}
according to its current  mixed strategy $\mathbf{q}_n(k)$, any active user $n$ in $\mathcal{B}(k)$ randomly chooses a channel ${a_n}(k)$ from its available channel set $\mathcal{A}_n$ in slot¡¡ $k$.

\textbf{2). Channel access and transmit:}
All the active users  transmit on the chosen channels, and they receive the instantaneous transmission throughput $R_n(k)$, which is determined by (\ref{eq:instantaneous_payoff}).

\textbf{3). Update mixed strategy:}
 All the active users $n\in\mathcal{B}(k)$ update their mixed strategies according to the following rules:
\begin{equation}
\label{eq:updating_rule}
\begin{array}{l}
 {q_{nm}}(k + 1) = {q_{nm}}(k) + b{{{r}}_{n}}(k)(1 - {q_{nm}}(k)),m = {a_n}(k) \\
 {q_{nm}}(k + 1) = {q_{nm}}(k) - b{{{r}}_{n}}(k){q_{nm}}(k),\;\;\;\;\;\;\;\;\;m \ne {a_n}(k), \\
 \end{array}
\end{equation}
where  $b$ is the learning step size, and ${{ { r}_n(k)}}$ is the normalized received payoff defined as follows:
\begin{equation}
\label{eq:normalized_payoff}
{{{r}}_{n}}(k) = \frac{R_n(k)}{R^{max}_n},
\end{equation}
where $R^{max}_n$ is the interference-free transmission throughput of user $n$ , i.e., $R^{max}_n=B \log \big(1+\frac{{P_{n}}d_{n}^{-\alpha }}{\sigma}\big)$.

All the inactive users keep their mixed strategies unchanged, i.e.,
\begin{equation}
\mathbf{q}_n(k+1)=\mathbf{q}_n(k), \forall n \in \mathcal{N}\backslash \mathcal{B}(k)
\end{equation}
\textbf{End loop}
\\
\rule{\linewidth}{1pt}

The rationale behind the update rule (\ref{eq:updating_rule}) is as follows: when a channel is selected and a positive payoff is received, the probability of choosing the channel in the next step increases while the probabilities of choosing other channels decrease accordingly. When all the players adhere to this rule, the system will finally converge to a stable state.
Note that the above learning algorithm is fully distributed since an active user only needs its individual action-payoff information. Furthermore, its asymptotical convergence property is characterized by the following theorem.

\begin{theorem}
\label{tm:learning_convergence}
When the learning step size  goes sufficiently small, i.e., $b \rightarrow 0$, the proposed distributed learning algorithm asymptotically  converges to a pure strategy NE point of robust spectrum access game  $\mathcal{F}_2$.
\end{theorem}

%
\begin{IEEEproof}
It has been rigorously proved that the stochastic learning automata converges to pure strategy Nash equilibria of exact potential games in  \cite{OSA_Xu_TWC12}. In methodology, the differences in this work are summarized as: i) all the users are always active in \cite{OSA_Xu_TWC12}, while they are randomly active or inactive in this work, and ii) for exact potential games, the change in the utility function cased by the unilateral action change of an arbitrary user is the same with that in the potential function, i.e.,
 \begin{equation}
 \label{eq:EPG_definition}
  \begin{array}{l}
    u_n(a_n,a_{-n})-u_n(a'_n,a_{-n})= \phi(a_n,a_{-n})-\phi(a'_n,a_{-n})
  \end{array}
 \end{equation}

When proving the convergence for exact potential games, the following inequality is vital (See equation (C.40) in \cite{OSA_Xu_TWC12}):
 \begin{equation}
 \label{eq:EPG_definition}
  \begin{array}{l}
 \big(u_n(a_n,a_{-n})-u_n(a'_n,a_{-n})\big)\big(\phi(a_n,a_{-n})-\phi(a'_n,a_{-n})\big)\ge0
  \end{array}
 \end{equation}
Note that the above inequality also holds for ordinal potential games (See equation (\ref{eq:potential_challenge3})). Thus, following similar lines for the proof given in \cite{OSA_Xu_TWC12} (Theorem 5), and with some additional modifications for processing the user active probability $\lambda_n$ \cite{Xu_TVT2015}, this theorem can be proved. However, to avoid unnecessary repetition,  the detailed proof is omitted.
\end{IEEEproof}




\section{Simulation Results and Discussion}

The simulation scenario follows the setting given in \cite{database_access1}. The cognitive APs are randomly located in a 500m$\times$500m square area. There are $M = 5$ channels with bandwidth $B=6$MHz, the noise power is $\sigma= -100$ dBm, and the path loss factor is $\alpha=4$. The distance  between AP $n$ and its associated boundary user is $d_n=20$ m. By inquiring the geo-location spectrum database, each AP operates with a specific transmission power $P_n$ and has a different set of available channels. The  step size for the  learning algorithm  is $b=0.1$.

\begin{figure}[!tb]
\centering
\includegraphics[width=3.35in]{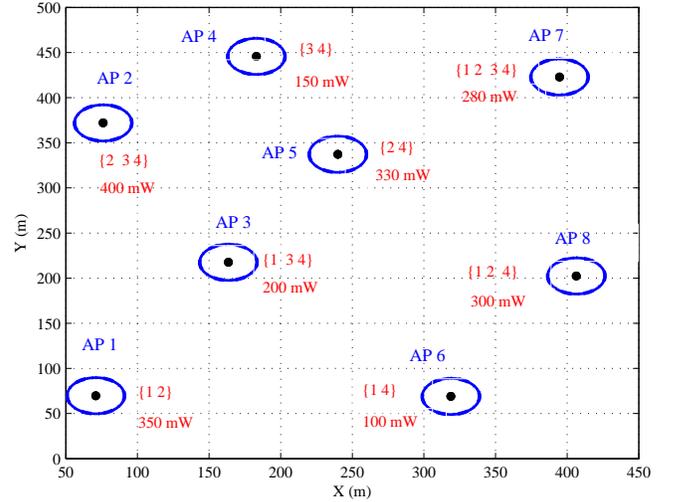}
\caption{A network consisting of eight cognitive APs. By inquiring the geo-location spectrum database, each AP knows
its available channel set and the transmitting  transmission power, e.g, the available channel set and transmission power of AP 1 are $\{1, 2\}$ and 350 mW, respectively.  }
\label{fig:specific_network}
\end{figure}

\subsection{Convergence behavior}
\begin{figure}[!tb]
\centering
\includegraphics[width=3.2in]{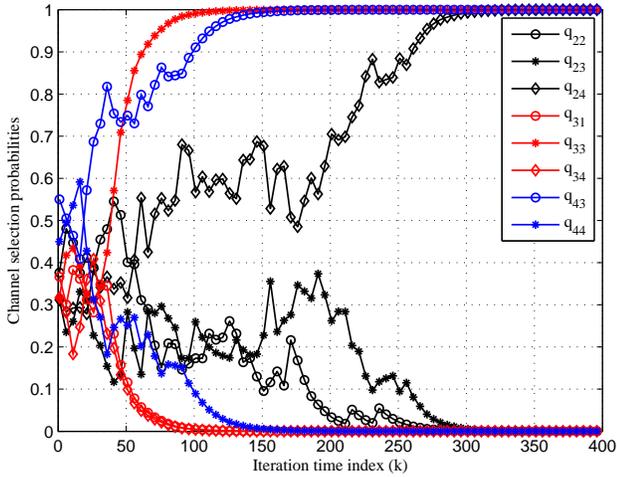}
\caption{The convergence of the proposed distributed learning algorithm.}
\label{fig:convergence}
\end{figure}

We consider a specific network which is shown in Fig. \ref{fig:specific_network}. Due to the different locations, the action (available channel) sets of the users are different. For example, the action set of users 1, 2, 3 and 4 are $\mathcal{A}_1=\{1,2\}$,
$\mathcal{A}_2=\{2,3,4\}$, $\mathcal{A}_3=\{1,3,4\}$ and $\mathcal{A}_4=\{3,4\}$ respectively. For presentation, it is assumed that all the users have the same active probability $\lambda=0.8$. The convergence behavior of the proposed distributed spectrum access  algorithm is shown in Fig. \ref{fig:convergence}. Specifically, the channel selection probabilities of users 2, 3 and 4 are presented. It is noted that the channel selection probabilities of user 2  converges $\{0, 0, 1\}$ over its available channel set $\{2, 3, 4\}$, which means that it finally chooses channel 4 for data transmission. Similarly, the selection probabilities of users 3 and 4 also finally converge. The results validate the convergence of the proposed distributed learning algorithm.

\subsection{Throughput performance}

For comparison, we consider four approaches: optimal, best NE, worst NE and the distributed learning algorithm. 1) Optimal: problem $\textbf{P2}$ is solved directly in a centralized manner. Due to the fact it is a combinatorial optimization problem, we apply the exhaustive search approach to obtain the optimal solution. 2) Best NE and worst NE: by assuming that information exchange among the users is available,
the best response algorithm \cite{Monderer96} is applied to achieve pure strategy NE of the robust spectrum access game $\mathcal{F}_2$ in a distributed manner. We carried out 500 independent trails and then take the best one and worst one respectively. Note that the best NE and worst NE serve as the upper and lower bounds of the game. 3) The proposed distributed learning: in the absence of information exchange and centralized controller, all the users adhere to the proposed distributed learning algorithm.

\begin{figure}[!tb]
\centering
\includegraphics[width=3.2in]{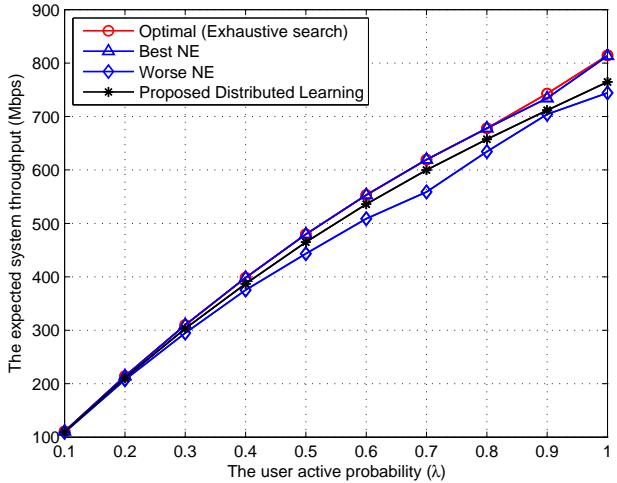}
\caption{The expected throughput  versus the active probabilities of the users.  }
\label{fig:throughput_probability}
\end{figure}

First, we evaluate the throughput performance for the specific network shown in Fig. \ref{fig:specific_network}. For presentation, all the users have the same active probability. The expected throughput when varying the user active probability is shown in Fig. \ref{fig:throughput_probability}. The result of the proposed learning is by simulating 200 independent trials and then taking the expected results. Some important results can be observed from the figure: i) the best NE is almost the same with the optimal solution, while the throughput gap between the worse NE and the optimal solution is trivial, which validate the effectiveness of the proposed robust spectrum game. ii) the proposed distributed learning is very close to the optimal one. Note that some similar numerical results on the throughput comparison were reported in \cite{Xu_magazine}.

\begin{figure}[!tb]
\centering
\includegraphics[width=3.2in]{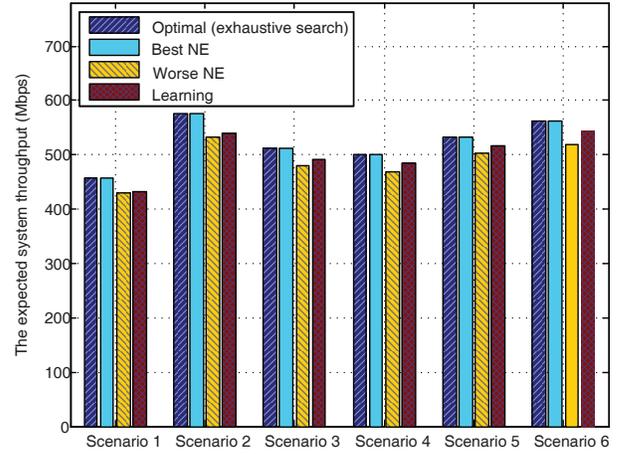}
\caption{The throughput performance comparison for six scenarios with heterogeneous active probabilities.}
\label{fig:throughput_heterogeneous}
\end{figure}

Secondly, we also consider the specific network shown in Fig. \ref{fig:specific_network}, but the active probabilities of the users are different. Specifically, the active probabilities of the users are set to
$\{0.1,0.2,0.3,0.4,0.5,0.6,0.7,0.8\}$ (tagged as Scenario 1),
$\{0.3, 0.3, 0.3, 0.6, 0.6, 0.9, 0.9, 0.9\}$ (tagged as Scenario 2),
$\{0.3, 0.4, 0.5, 0.5, 0.5, 0.6, 0.7, 0.8\}$ (tagged as Scenario 3),
$\{0.4, 0.4, 0.4, 0.6, 0.6, 0.6, 0.6, 0.6\}$ (tagged as Scenario 4),
$\{0.3, 0.6, 0.6, 0.6, 0.6, 0.6, 0.6, 0.7\}$ (tagged as Scenario 5),
$\{0.6, 0.6, 0.6, 0.6, 0.6, 0.6, 0.6, 0.7\}$ (tagged as Scenario 6), respectively.
The throughput comparison results are shown in Fig. \ref{fig:throughput_heterogeneous}. It also noted from the figure that
for the scenarios with heterogeneous active probabilities, the best NE is almost the same with the optimal solution and the proposed distributed learning is close to the optimal one. These results validate the effectiveness of the formulated spectrum access game as well as the proposed distributed learning algorithm, in both homogeneous  and heterogeneous scenarios.

\begin{figure}[!tb]
\centering
\includegraphics[width=3.35in]{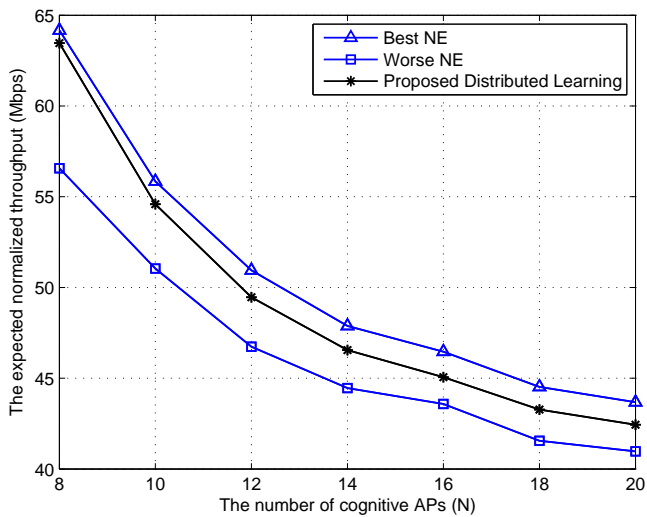}
\caption{The throughput performance comparison for general networks (the active probabilities of all the users are $\lambda=0.8$). }
\label{fig:throughput_number}
\end{figure}

Thirdly, we evaluate the throughput performance for general networks. Specifically, the cognitive APs are randomly located in the square region. Each channel is independently vacant with probability $\theta=0.7$ for each  AP, and the transmission power of each AP is randomly chosen from the set $\{100, 200, 250, 300, 350, 280, 400\}$mW. The throughput comparison when varying the number of cognitive APs is shown in Fig. \ref{fig:throughput_number}. For each number of APs, e.g., $N=10$, we simulate $100$ independent topologies and take the average result. When the network scales up, exhaustive search is not feasible due to the heavy computational complexity. However, it is believed that the best NE would be very close to the optimal one. It is shown from the figure that the proposed distributed learning is close to the best NE, which again validates the proposed game-theoretic solution. Also, as the network scales up, the normalized throughput decreases due to the increase in the mutual interference, as can be expected.

To summarize, the simulation results show that the best NE is almost the same with optimal solution, and the proposed distributed learning algorithm is very close to the optimal one. Recalling the dynamic and incomplete information constraints in the considered system, we believe that the proposed game-theoretic learning solution is desirable for practical applications.

%

\section{Conclusion}
This paper investigated the problem of database-assisted spectrum access in dynamic networks, in which the active user set is varying. Since there is no central controller, it encounters dynamic and incomplete information constraints.
To solve this challenge, we formulated a state-based spectrum access game and a robust spectrum access game.
We proved that the two games are ordinal potential games with the (expected) aggregate interference serving as the potential functions, and proposed a distributed learning algorithm without information exchange to achieve the pure strategy Nash equilibrium (NE) of the games. Simulation results show that the best NE is almost the same with the optimal solution and the achievable throughput of the proposed learning algorithm is very close to the optimal one, which  validates the effectiveness of the proposed game-theoretic solution. Note that there are still some new challenges and open issues to be studied. For example,
users can access more than one channel when equipped with the multiple radio technology. A game-theoretic carrier aggregation in unlicensed spectrum bands is ongoing and will be reported soon.



\


%

\ifCLASSOPTIONcaptionsoff
  \newpage
\fi



%
\bibliographystyle{IEEEtran}
\bibliography{IEEEabrv,reference}

%

%








\end{document}